\documentclass[twocolumn,showpacs,preprintnumbers,amsmath,amssymb]{revtex4}

\usepackage{graphicx}
\usepackage{dcolumn}
\usepackage{bm}

\begin{document}

\preprint{APS/123-QED}

\title{A Simple Probabilistic Algorithm for Detecting Community
Structure in Social Networks}

\author{Wei Ren}
 \email{renwei@amss.ac.cn}
\author{Guiying Yan}
 \author{Xiaoping Liao}
 \author{Lan Xiao}
 \affiliation{Academy of Mathematics and Systems Science \\
 Chinese Academy of Science}

\begin{abstract}
With the growing number of available social and biological networks,
the problem of detecting network community structure is becoming
more and more important which acts as the first step to analyze
these data. The community structure is generally regarded as that
nodes in the same community tend to have more edges and less if they
are in different communities. We come up with \textbf{SPAEM}, a
\textbf{S}imple \textbf{P}robabilistic \textbf{A}lgorithm for
detecting community structure which employs
\textbf{E-M}(Expectation-Maximization) algorithm. We also give a
criterion based on minimum description length to identify the
optimal number of communities. \textbf{SPAEM} can detect overlapping
nodes and handle weighted networks. It turns out to be powerful and
effective by testing simulation data and some widely known data
sets.
\end{abstract}


\maketitle

\section{INTRODUCTION}

Many systems can be represented by networks where nodes denote
entities and links denote existing relation between nodes, such
systems may include the web networks \cite{socialnetwork}, the
biological networks \cite{biologynetwork}, ecological web
\cite{dolphin} and social organization networks \cite{zachary}. Many
interesting properties have also been identified in these networks
such as small world\cite{SmallWorld} and power law distribution
\cite{scalefree}, one property that attracts much attention is the
network community structure, which is the phenomenon that nodes
within the same community are more densely connected than those in
different communities \cite{GN}. It is important in the sense that
we can get a better understanding about the network structure.

This problem has been studied by researchers from different
perspectives. Earlier approaches for identifying communities could
be divided into two categories: the hierarchical approach and
divisive approach. The former merged two closest nodes into one
community recursively until the whole network became one single
community and the latter worked from the top to bottom which split
the whole network into 2 communities recursively until every node
was a community. These algorithms usually needed a measure to
evaluate the closeness or dissimilarity between two nodes, see
\cite{GN,GNbio,haijun,Italian,shihuakernel}.

An important modularity measure for evaluating the goodness of
community structure was proposed by Newman \cite{modularity} and
several algorithms worked by maximizing it
\cite{modularitymatrix,newmaneigen,spectral,extreamoptimization}.
This measure was very efficient in  characterizing community
structure for networks with balanced structure, however, the
internal scale problem in its definition \cite{limit} made it fail
to work well for unbalanced networks such as those whose communities
varied in size and degree sequence. Quite recently, an information
based algorithm by Martin \cite{info} could accurately resolve
communities and in particular can to some extent get over the scale
problem of modularity.

Also, researchers \cite{natureoverlap} found that communities were
overlapping rather than disjoint, subsequent algorithms
\cite{shihuafuzzy,info,shihuanmf} were designed to deal with
overlapping communities. A mixture model by Newman \cite{newmanEM}
could automatically detect patterns inside a network, meanwhile, it
was able to detect overlapping nodes as a byproduct.

All these state-of-art algorithms motivate us to treat the community
detection problem as a probabilistic inference problem, we should
mine the internal information which determines the network topology.
These internal information gives insight to the network structure.
Our work is inspired by probabilistic latent semantic analysis
\cite{plsa}  which is a powerful algorithm in text ming, it models
that a term occurs in a document if they are under the same latent
topic. This idea is employed here to detect community structure in
complex networks.

\section{METHOD}

\begin{figure}
\begin{center}
\includegraphics[scale=0.5]{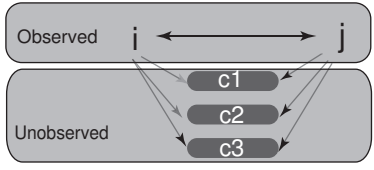}
\caption{\label{fig:model}An observed edge exists between nodes
$i,j$ due to the fact that they participate in the same community,
which is hidden or unobserved information. Our model mines these
hidden information such that they generate the network with the
highest probability.}
\end{center}
\end{figure}

 Assume that the network considered is undirected and
unweighted with $n$ nodes, let $A$ denote the adjacent matrix and
$N(i)$ the neighbors of node $i$. Assume that two nodes $i$ and $j$
have an edge if they belong to the same community, which is hidden
information, see FIG \ref{fig:model}. According to the model,
community membership $g_{ij}$ can be assigned to edge $e_{ij}$, such
that $g_{ij}=r$ if and only if two nodes of edge $e_{ij}$ belong to
community $r$.

Suppose $c$ communities are to be detected, let $\pi_r$ be the
probability of community $r$, which can be viewed as the fraction of
nodes in community $r, r=1,2,...,c$. The conditional probability
$Pr(i|r)$ of node $i$ appearing in community $r$, denoted by
$\beta_{r,i}$, satisfies $\sum_{i=1}^n \beta_{r,i}=1$. In fact
$\beta_{r,i}$ can be  viewed as the importance of node $i$ in
community $r$, the larger the value of $\beta_{r,i}$ is, the more
important node $i$ is in community $r$. Let $\pi$ and $\beta$ denote
set of parameters $\{\pi_r,r=1,2,...,c\}$ and
$\{\beta_{r,i},r=1,2,..,c, i=1,2,...,n\}$, respectively.

Naturally, the probability of edge $e_{ij}$ existing and belonging
to community $r$ is modeled as
\[Pr(e_{ij}, g_{ij}=r|\pi,\beta)=\pi_r\beta_{r,i}\beta_{r,j}\]
In fact, $Pr(e_{ij},g_{ij}=r|\pi,\beta)$ can be viewed as the
contribution of community $r$ to the formation of edge $e_{ij}$.
Then probability $Pr(e_{ij}|\pi,\beta)$ of  the existence of edge
$e_{i,j}$ is the sum of contribution from all communities
$r=1,2,..,c$, namely
\[Pr(e_{ij}|\pi,\beta)= \sum_{r=1}^c  Pr(e_{ij},g_{ij}=r|\pi,\beta)=\sum_{r=1}^c \pi_r
\beta_{r,i}\beta_{r,j}\] The observed information is the edge
$e_{ij}$, however, it's determined by the unobserved parameters
$\pi, \beta$, so these unobserved parameters $\pi,\beta$ determine
the network topology $A$, this is exactly the idea depicted in FIG
\ref{fig:model}.

Next, the log probability of network $A$ under parameters
$\pi,\beta$ can be modeled as
\begin{eqnarray}
\label{eqn:LL} \nonumber LL&=\log Pr(A|\pi,\beta)= \sum_{i=1}^n
\sum_{j:j\in N(i)} \log Pr(e_{ij}|\pi,\beta)
\\
&=\sum_{i=1}^n\sum_{j:j\in N(i)}\log (\sum_{r=1}^c
\pi_r\beta_{r,i}\beta_{r,j})
\end{eqnarray}
Parameters $\pi,\beta$ should be estimated to maximize
Eq(\ref{eqn:LL}). However, $LL$ in Eq(\ref{eqn:LL}) contains log of
sums and is difficult to optimize but can be optimized easily by
Expectation-Maximization algorithm.

\subsection{The EM Formula}
The EM algorithm is  proposed to maximize probability that contains
latent variables \cite{EMfirst}, it computes the posterior
probability of the latent variables under the observed data and
currently estimated parameters in the E-step and updates parameters
with these posterior probabilities in the M-step. The posterior
probability $Pr(g_{ij}=r |A,\pi,\beta)$ of edge $e_{i,j}$ belonging
to community $r$ under the observed network $A$ and parameter
$\pi,\theta$, denote this probability by $q_{ij,r}$, then
\begin{eqnarray}\nonumber q_{ij,r}=Pr(g_{ij}=r|A,\pi,\beta)=\frac{Pr(g_{ij}=r,
A|\pi,\beta)}{Pr(A|\pi,\beta)}
\\
\nonumber=\frac{Pr(e_{ij},g_{ij}=r,A|\pi,\beta)}{Pr(A|\pi,\beta)}
\end{eqnarray}
by simple deduction the E-step formula can be obtained:
\begin{eqnarray}\label{eqn:Estep}q_{ij,r}=Pr(g_{ij}=r|A,\pi,\beta)=\frac{\pi_r
\beta_{r,i}\beta_{r,j}}{\sum_{s=1}^c \pi_s \beta_{s,i}
\beta_{s,j}}\end{eqnarray} In fact, $q_{ij,r}$ is the fraction of
contribution from community $r$ under the observed matrix $A$ and
parameters $\pi,\beta$. Obviously, the expected log-probability of
the network is
\begin{eqnarray}\label{LLL}\nonumber \overrightarrow{LL}=\sum_{i=1}^n\sum_{j:j\in N(i)} \sum_{r=1}^c q_{ij,r}\ln Pr(e_{i,j},g_{ij}=r|\pi,\beta)\\
= \sum_{i=1}^n \sum_{j:j\in N(i)} \sum_{r=1}^c q_{ij,r} \ln (\pi_r
\beta_{r,i} \beta_{r,j})\end{eqnarray}

Combining with the constraints that $\sum_r \pi_r=1, \sum_{i=1}^n
\beta_{r,i}=1, r=1,2,..,c$, the lagrange form of
$\overrightarrow{LL}$ is:
\begin{eqnarray}\label{lagrange}
L= \nonumber\sum_{i=1}^n \sum_{j:j\in N(i)} \sum_{r=1}^c q_{ij,r}
\ln (\pi_r \beta_{r,i} \beta_{r,j}) \\
 +\alpha(\sum_{r=1}^c \pi_r -1) +\sum_{r=1}^c \gamma_r
(\sum_{i=1}^n \beta_{r,i}-1)
\end{eqnarray}
where $\alpha, \gamma_r,r=1,2,...,c$ are lagrange multipliers. The
derivatives of $L$ in Eq(\ref{lagrange}) are:
\begin{eqnarray}
\label{derivative1}\frac{\partial L}{\partial \pi_r}=\sum_{i=1}^n
\sum_{j:j\in N(i)}
q_{ij,r}+\alpha \\
\label{derivative2}
 \frac{\partial L}{\partial \beta_{r,i}}=\sum_{j:j\in N(i)}
q_{ij,r}+\gamma_r
\end{eqnarray}
By setting the derivative in
Eq(\ref{derivative1}),Eq(\ref{derivative2}) to zero and combining
the constraints $\sum_r \pi_r=1, \sum_{i=1}^n \beta_{r,i}=1,
r=1,2,..,c$, the M-step formulas are:
\begin{align}
\pi_r& = \frac{\sum_i\sum_{j:j\in N(i)} q_{ij,r}}{\sum_i \sum_{j:\in
N(i)} \sum_{s=1}^c q_{ij,s}}
\\
\beta_{r,i}&=\frac{\sum_{j:j\in N(i)} q_{ij,r}}{\sum_{k=1}^n
\sum_{j:j\in N(k)} q_{kj,r}}
\end{align}

In the E-step, the membership of an edge is influenced by its nodes
while in the M-step, the node importance in communities  is
influenced by the membership of all its links. By iterating E-steps
and M-steps, $LL$ in Eq(\ref{eqn:LL}) will increase.

Once all the parameters are estimated, the preference of node $i$
belonging to community $s$ is computed as $u_{s,i}=\pi_s
\beta_{s,i}$, and node $i$ is assigned to community $r$ such that
$r=argmax_{s}\{u_{s,i}=\pi_s \beta_{s,i},s=1,2,..,r\}.$ $u_{s,i}$s
can be normalized so that their sum is 1 to comply with probability
normalization condition . In fact, this gives a soft assignment and
can be used to detect overlapping nodes. Suppose for node $i$,
$r=argmax_{s}\{u_{s,i}=\pi_s \beta_{s,i},s=1,2,..,r\}$, empirically
node $i$ is an overlapping node if there is another community $s$
such that $\frac{u_{s,i}}{u_{r,i}}>1/10$.

 Parameters $\pi,\beta$ are
initialized with random values and iterated using E-step and M-step
until $LL$ stabilizes. To avoid the algorithm getting stuck in a
local maxima, we adopt restart strategy which runs the EM algorithm
several times with different initial parameter values.

Suppose the network has totally $l$ edges, obvious the algorithm has
a linear time complexity $O(cl)$, which makes it an appealing
approach for detecting large scale networks. Note that the actual
running time is also relevant to the number of EM iterations and the
number of restarts. We name our model \textbf{SPAEM} for easier
representation, abbreviation for \textbf{S}imple
\textbf{P}robabilistic \textbf{A}lgorithm which employs the idea of
\textbf{E}xpectation and \textbf{M}aximization framework.
\subsection{Model Selection Issue}
\textbf{SPAEM} needs a pre-specified community number $c$ and this
is regarded as \emph{prior} knowledge. However, the determination of
$c$ is a non-trivial task and is difficult when no prior knowledge
can be obtained.  We try to handle it using Minimum Description
Length principle \cite{mdl}.

In general, $LL$ in Eq(\ref{eqn:LL}) increases as $c$ increases,
meanwhile, an extra cost has to be paid for due to the increase in
the number of parameters $K=(c-1)+c(n-1)$. There should be some
balance between $LL$ and $K$, and the idea of minimum description
length principle can be employed here \cite{mdl}. According to this
principle, the code length needed to describe the network data is
composed of 2 parts whereas the first part describes the coding
length of the network using \textbf{SPAEM}  while the second part
gives the length for coding all parameters of \textbf{SPAEM} itself.
The length needed for the coding network using \textbf{SPAEM} is
obviously $-LL /2$ (note that every edge is added twice). To code
the parameters, a precision $\epsilon$ has to be pre-specified. With
this precision $\epsilon$, parameters smaller than $\epsilon$ are
not coded and get a description length of 0, otherwise coding the
parameter $\pi_r$ needs length $\log(\frac{\pi_r}{\epsilon})$ and
$\beta_{r,i}$ needs length $\log(\frac{\beta_{r,i}}{\epsilon})$, so
the total length $H$ for coding the model is
\begin{eqnarray}\label{MDL}
\nonumber H=-LL/2+\sum_{r=1}^c \log(\frac{\pi_r}{\epsilon})I(\pi_r
\geq
\epsilon)+ \\
\sum_{r=1}^c
\sum_{i=1}^n\log(\frac{\beta_{r,i}}{\epsilon})I(\beta_{r,i}\geq
\epsilon)
\end{eqnarray}

Value $c$ should be chosen as the one  which generates the minimum
description length $H$ in Eq(\ref{MDL}). Choosing precision
$\epsilon$ is tricky but very important in Eq(\ref{MDL}). Smaller
$\epsilon$ may cause longer code for parameters and hence will
always prefer models with small $c$. In fact, it's shown that
networks are organized in a hierarchical way \cite{hier}, the choice
$\epsilon$ gives a lever for viewing networks in different
resolutions. It's intuitively clear that $\epsilon$ should be on the
scale of $1/n$ due to the normalization condition $\sum_{i=1}^n
\beta_{r,i}=1$. Typically, if node $i$ belongs to community $r$,
$\beta_{r,i}$ will be on the scale of $1/n$ and be much smaller than
$1/n$ if not belongs to this community. Here $\epsilon$ is set to
$1/(3n)$. This precision is totally empirical but as will be shown
in next section that for well clustered networks, the model
selection results are robust to the choice of $\epsilon$ ranging
from $1/n$ to $1/(7n)$.

\section{EXPERIMENT}
\subsection{Zachary Club Network}
\begin{figure}
\begin{center}
\includegraphics[scale=0.35]{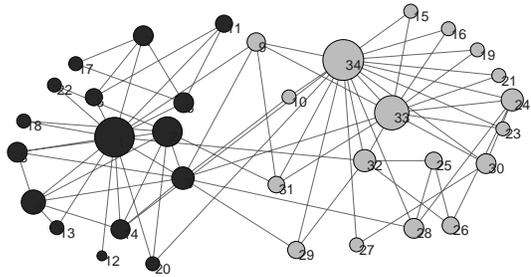}
\caption{\label{zachary}Zachary club network:Node color indicates
community and node size indicates $u_{r,i}$. Clearly node 9,10,31
are overlapping nodes and have been identified by our algorithm.}
\end{center}
\end{figure}

 The  famous zachary club network is about acquaintance
relationship between 34 members \cite{zachary}. The club splits to 2
parts due an internal dispute so it naturally has community
structure. By setting $c=2$, we run our algorithm and get exactly
the original 2 communities, see FIG \ref{zachary}. Node color
indicates community and node size indicates the value of
$u_{r,i}=\pi_r\beta_{r,i}$ which can partially measures the
importance of node $i$ in community $r$. Node 1,2,33,34 are
important nodes found by \textbf{SPAEM} and can be verified
intuitively from the network.

\textbf{SPAEM} gives soft assignment to each node so is capable of
detecting overlapping nodes, see Table \ref{table:zachary}.  To
compare the ability in detecting overlapping nodes, we also include
$q_{ir}$ used to assign communities in the mixture model
\cite{newmanEM}. Clearly, nodes 1,2,33,34 are not overlapping nodes,
but node 9 is. The mixture model also can detect this, however, by
checking corresponding probabilities, see Table \ref{table:zachary},
\textbf{SPAEM} shows more accuracy revealing the extent of
overlapping.
\begin{table}
\caption{\label{table:zachary}Result on zachary network. $u_{r,i}$
is calculated by $u_{r,i}=\pi_r*\beta_{r,i}$, which is interpreted
as the preference of node $i$ belonging to community $r$. The
$q_{ir}$s in the mixture mode \cite{newmanEM} are also included, to
facilitate comparison, we normalize $u_{s,i}$ so they add up to 1. }
\begin{ruledtabular}
\begin{tabular}{|l|c|c|c|r}
\hline Node ID & $u_{1,i}$ & $u_{2,i}$ & $\frac{u_{1,i}}{u_{1,i}+u_{2,i}}$ & $q_{i1}$ \footnotemark[1]\\
\hline
1 &  3.30E-05   & 0.1025  & 0.00 & 0.00\\

2 &  4.86E-06   & 0.0577  & 0.00 & 0.00 \\

9   & 0.0219 & 0.0101 & 0.69 & 0.96 \\

13 & 5.83E-36   & 0.0128 & 0.00 & 0.00\\
31 & 0.0179     & 0.0078 & 0.70 & 0.92\\
33 & 0.0769 &  1.55E-08 & 1.00 & 1.00\\
34 & 0.1090  & 8.20E-06 &1.00 & 1.00\\
\end{tabular}
\footnote[1]{$q_{ir}$ is defined in \cite{newmanEM} as the
probability of node $i$ belonging to community $r$.}
\end{ruledtabular}
\end{table}

\subsection{American College Football Team Network}
The second network investigated is the college football network
which represents the game schedule of the 2000 season of Division I
of the US college football league \cite{GN}. The nodes in the
network represent the 115 teams, while the links represent 613 games
played. The teams are divided into 12 conferences and generally
games are more frequent between members of the same conference than
between teams of different conferences.
\begin{figure}
\begin{center}
\includegraphics[scale=0.5]{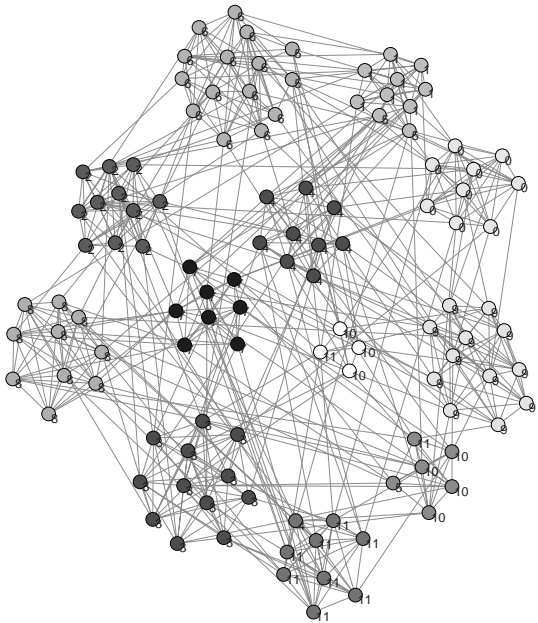}
\caption{\label{fig:football}Result of \textbf{SPAEM} for American
football network: Node label indicates real community membership.
Nodes belonging to the same community detected by  \textbf{SPAEM}
are placed adjacently.}
\end{center}
\end{figure}

\begin{figure}
\begin{center}
\includegraphics[scale=0.5]{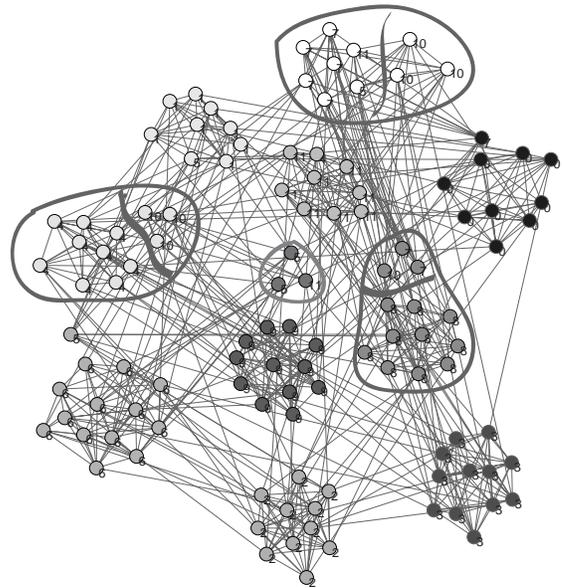}
\caption{\label{fig:footballnewman}Result of the mixture model
\cite{newmanEM} for American football network: Node label indicates
real community membership. Nodes belonging to the same group
detected are placed together, groups which are not the common sense
community structure are marked using cycled line. Some of these
groups are formed by nodes from two real communities. Also there is
a 3 node group which is clearly not a community. }
\end{center}
\end{figure}
The result of \textbf{SPAEM} and the mixture model \cite{newmanEM}
is depicted in FIG \ref{fig:football} and FIG
\ref{fig:footballnewman}, respectively. \textbf{SPAEM} basically
uncovers the original community structure. However, the mixture
model gets a very different result,  see FIG
\ref{fig:footballnewman}. This is because the group it detected is a
set of nodes with similar linkage property so may not be common
sense community. The 3 node group in the middle of FIG
\ref{fig:footballnewman} is obviously not a community. There are
still other groups consisting of nodes from different communities,
see FIG \ref{fig:footballnewman}. The mixture model can detect
patterns but it can not differentiate different kinds of patterns,
in other words, it can not tell whether a detected group is a
community.
\subsection{Comparison With Other Methods \label{subsection:test}} A modularity measure $ Q=\sum_{r=1}^c (\frac{l_{rr}}{l}
-(\frac{d_r}{2l})^2)$ is proposed by Newman \cite{modularity}, where
$l_{rr}$ is the number of links in community $r$, $d_r$ is the total
degree in community $r$, $l$ is the total number of edges in the
network . Good community structure usually indicates  large value of
$Q$. But there is a scale $l$ in the definition of $Q$ and this may
cause problem in some networks \cite{limit,info}. Such networks
include those whose communities vary in size and degree sequence.

\begin{figure}
\includegraphics[scale=0.4]{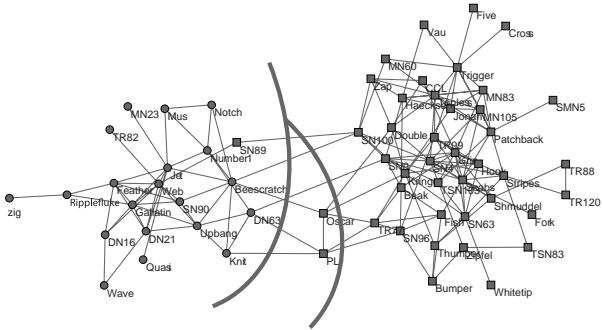}
\caption{\label{fig:dolphin}Dolphin network: Node shape denotes the
real split. The grey line shows the result by \textbf{SPAEM} with
only 1 mistake, the left and right black line indicates the results
of algorithms in \cite{spectral}. }
\end{figure}

Dolphin social network reported by Lusseau \cite{dolphin} provides a
natural example where communities vary in size. In this network, two
dolphins have a link with each other if they are observed together
more often than expected by chance. The original two communities
have different sizes, with one containing 22 dolphins and the other
40. Setting $c=2$, \textbf{SPAEM} only misclassifies one node and
gets exactly the same result as the GN algorithm  \cite{GN} and the
information based algorithm \cite{info}, however, the modularity
based method \cite{spectral} gets different result, as depicted in
FIG \ref{fig:dolphin}.

It is shown that the modularity algorithm works well for networks
whose communities roughly have the same size and degree sequence,
but may not provide very competitive results when the communities
differ in size and degree sequence \cite{info}. To show the way
\textbf{SPAEM} handles these situations, we conduct the same 3 sets
of test as done in \cite{info}: symmetric, node asymmetric, link
asymmetric. In the symmetric test, each network is composed of 4
communities with 32 nodes each, each node has an average degree of
16, $k_{out}$ is the average number of edges linking to nodes in
different communities. In the node asymmetric test, each network is
composed of 2 communities with 96 and 32 nodes respectively,
$k_{out}$ has the same meaning as in the symmetric test. $k_{out}$
is set to 6,7,8 in both the symmetric and node asymmetric case, as
$k_{out}$ increases, it becomes difficult to detect real community
structure. In the link asymmetric test, 2 communities each with 64
nodes differ in their average degree sequence, nodes in one
community have average 24 edges and in the other community have only
8 edges, setting $k_{out}=2,3,4$. Table \ref{table:benchmark} gives
the results of our algorithm compared to other algorithms
\cite{info,modularity,newmanEM}. Note that the results of the
information algorithm and the modularity algorithm are cited from
\cite{info} while results of the mixture model are calculated by the
authors. We have to admit that the information algorithm outperforms
all other 3 algorithms, especially in the node asymmetric and link
asymmetric tests. \textbf{SPAEM} outperforms the modularity
algorithm \cite{modularity} in the symmetric and node asymmetric
tests. The mixture model \cite{newmanEM} seems to perform not so
well in the symmetric test, this might be due to that the groups it
discovers may not be communities due to fuzzy structure of these
networks as $k_{out}$ increases.
\begin{table}
\caption{\label{table:benchmark}Results on the benchmark test on 3
experiment: symmetric, node asymmetric, link asymmetric.}
\begin{ruledtabular}
\begin{tabular}{lllccc}
Test & $k_{out}$ &SPAEM & Compression\footnotemark[1] &
Modularity\footnotemark[2] & Mixture\footnotemark[3]
\\
 Symmetric  & 6 &  0.99   & 0.99 & 0.99 & 0.92\\
& 7 &  0.95   & 0.97 & 0.97 & 0.81\\
& 8& 0.84 &  0.87  & 0.89 & 0.64\\
Node  & 6 & 0.97 & 0.99 & 0.85 & 0.97 \\
Asymmetric& 7 & 0.92 & 0.96 & 0.80 & 0.92 \\
& 8 & 0.79 & 0.82 & 0.74 & 0.74\\
Link  & 2 & 0.98 & 1.00 & 1.00 & 0.99\\
Asymmetric& 3 & 0.94 & 1.00 & 0.96 & 0.94\\
& 4 & 0.84 & 1.00 & 0.74& 0.70\\
\end{tabular}
\end{ruledtabular}
 \footnotetext[1]{Information method in\cite{info}}
 \footnotetext[2] {Modularity based method in\cite{modularity}}
 \footnotetext[3]{Mixture model in \cite{newmanEM}}
\end{table}
\subsection{Handling Weighted Network}

\textbf{SPAEM} can also be extended to handle weighted networks.
  Suppose the weighted adjacent matrix of the
network is $W_{n\times n}$ with its entries
$w_{i,j},i=1,2,...,n,j=1,2,...,n$, then the loglikelihood of the
network becomes
\begin{eqnarray}
LL=\sum_{i=1}^n\sum_{j:j\in N(i)} w_{i,j}\log (\sum_r
\pi_r\beta_{r,i}\beta_{r,j})
\end{eqnarray}
$\overrightarrow{LL}$ becomes
\nonumber \begin{eqnarray}\overrightarrow{LL}=\sum_{i=1}^n \sum_{j:j\in N(i)} \sum_r w_{i,j}q_{ij,r}\ln Pr(e_{i,j}\in
r)\\
\nonumber = \sum_{i=1}^n \sum_{j:j\in N(i)} \sum_r w_{i,j}q_{ij,r} \ln (\pi_r \beta_{r,i}
\beta_{r,j})\end{eqnarray} The E-step is unchanged but M-step becomes
\begin{eqnarray}
\nonumber \pi_r& = \frac{\sum_i\sum_{j:j\in N(i)} w_{i,j}q_{ij,r}}{\sum_i \sum_{j:j\in N(i)} \sum_s w_{i,j}q_{ij,s}}
\\
\nonumber \beta_{r,i}&=\frac{\sum_{j:j\in N(i)} w_{i,j}q_{ij,r}}{\sum_{k=1}^N
\sum_{j:j\in N(k)} w_{k,j}q_{kj,r}}
\end{eqnarray}

Intuitively the M-step formula is reasonable since links with
greater weights contribute more to corresponding parameters.

 To test \textbf{SPAEM} on weighted networks, simulation test is done as that in \cite{weighted}.
 This set of test is based on the above symmetric test when
 $k_{out}=8$:
 For each of the 100 networks in the Symmetric Test with $k_{out}=8$,  the weight of edges within a certain
community is raised to $w=1.4,1.6,1.8,2$, while the weight of edges
running between communities is unchanged(with weight 1). As weight
$w$ increases from 1.4 to 2, models should improve their power in
detecting community structure. Results of \textbf{SPAEM} are shown
in Table \ref{table:weighted} as well as the results in
\cite{weighted} for comparison(note that the results in
\cite{weighted} are directly cited rather than recalculated).
\textbf{SPAEM} generally outperforms the model in \cite{weighted}.

\begin{figure}
\begin{center}
\includegraphics[scale=0.5,bb=0 340 520 500]{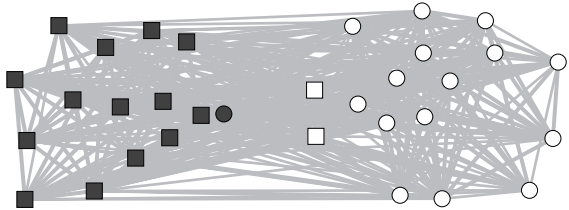}
\caption{\label{fig:weighted}Results on the simulated weighted
network. Node shape shows the original community while node color
indicates the community structure detected by \textbf{SPAEM}.}
\end{center}
\end{figure}

\begin{table}
\caption{\label{table:weighted} Benchmark test on weighted network
designed by \cite{weighted}. There are 4 communities each with 32
nodes in the network with $k_{out}=8$. As $w$ increase from 1.4 to
2, both methods respond positively but \textbf{SPAEM} gets better
results.}
\begin{ruledtabular}
\begin{tabular}{lcc}

 & SPAEM  & Markov \footnotemark[1]
\\
\hline
 $w=1.4$ & 0.96 & 0.89
 \\
 $w=1.6$ & 0.98 & 0.94
 \\
 $w=1.8$ & 0.99 & 0.97
 \\
 $w=2$ & 0.99 & 0.98
 \\
 \hline
\end{tabular}
\end{ruledtabular}
\footnotetext[1]{Random walk model in \cite{weighted}}
\end{table}
The limitation with the above simulation test is that any algorithm
will respond positively when $w$ increases and that the original
unweighted networks already have clear community structure. Now we
devise a more elaborate example: consider a network with 32 nodes,
each node pair has an edge with probability $p_{rand}$, obviously,
this network has no community structure. Let node 1 to 16 be in
group 1, node 17 to 32 be in group 2. Weight of edges inside each
group is raised to 1.5 with probability $p_{weight}$ but weight of
edges running between groups is unchanged. Now the only thing that
can differentiate these two groups is the weight of edges. By
setting $p_{rand}=0.8$ and $p_{weight}=0.8$, \textbf{SPAEM} uncovers
the two groups with only 3 mistakes, see FIG \ref{fig:weighted}.
This shows \textbf{SPAEM} is able to take good use of edge weight.
\subsection{Model Selection Test}
\begin{figure}
\includegraphics[scale=0.5]{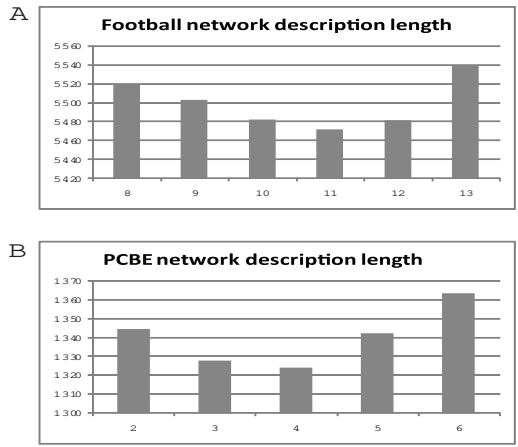}
\caption{\label{fig:modelselection}(a): Model selection result for
American football team network. (b): Model selection result for the
 the journal citation network.}
\end{figure}

 Now, the minimum
description length $H$ defined in Eq(\ref{MDL}) is employed for
\textbf{SPAEM} to select $c$, the optimal number of communities, and
the precision is empirically set to $1/3n$.  The criterion indicates
that 11 communities in the American football network\cite{GN} should
be detected, see FIG \ref{fig:modelselection}.a,  the result seems
to be wrong since there should be 12 communities, however, there is
a conference ``Independents'' which can not be a really conference
because teams in it play games with adjacent conferences. This
criterion also determines 4 communities in the journal citation
network, see FIG \ref{fig:modelselection}.b. These two results shows
$H$ in Eq(\ref{MDL}) and precision $1/3n$ are sound.

To further test the validity of the model selection principle ,
model selection results on the above simulation experiments
(Symmetric, Node Asymmetric, Link Asymmetric) are presented in Table
\ref{table:modelselection}. Combined with the model selection
principle, \textbf{SPAEM} gives very competitive results in all
these three tests. One weird thing is that in the node asymmetric
case, the accuracy of \textbf{SPAEM} increases as $k_{out}$
increases, this is partly because that the penalty term for
describing the model parameters in Eq.(\ref{MDL}) favors small
number communities, this also in turn verifies that selection
criterion and the precision is reasonable.
\begin{table}
\caption{\label{table:modelselection}Model selection result: Each
entry is the fraction of networks identified with the correct number
of communities, the number in the parentheses indicates the average
number of communities identified by the corresponding algorithm. }
\begin{ruledtabular}
\begin{tabular}{lllcc} Test & $k_{out}$
&SPAEM-MDL& Information\footnotemark[1] & Modularity\footnotemark[2]
\\
 Symmetric  & 6 & 1.00(4.00)    & 1.00(4.00) & 1.00(4.00)\\
& 7 &  1.00(4.00)   & 1.00(4.00) & 1.00(4.00)\\
& 8& 0.65(3.60) &  0.14(1.93) & 0.70(4.33)\\
Node & 6 & 0.82(2.18) & 1.00(2.00) & 0.00(4.95) \\
Asymmetric & 7 & 0.83(2.17) & 0.80(1.80)& 0.00(4.97) \\
& 8 & 0.93(2.07) & 0.06(1.06) & 0.00(5.29)\\
Link  & 2 & 1.00(2.00) & 1.00(2.00) & 0.00(3.10)\\
Asymmetric& 3 & 1.00(2.00) & 1.00(2.00)  &0.00(4.48)\\
& 4 & 1.00(2.00) & 1.00(2.00)  &0.00(5.55)\\
\end{tabular}
\end{ruledtabular}
 \footnotetext[1]{Information method \cite{info}}
 \footnotetext[2] {Modularity method \cite{modularity}}
\end{table}
\subsection{Model Selection Discussion}
The model selection criterion in Eq(\ref{MDL}) is sensitive to the
choice of the accuracy $\epsilon$, different $\epsilon$ would lead
to different model selection results. Intuitively, small $\epsilon$
will favor smaller number of communities and large $\epsilon$ tends
to identify large number of communities. In fact, it's shown that
complex networks may be organized in the hierarchical structure
which allows us to view them in different resolutions \cite{hier}.
The accuracy  $\epsilon$ indeed provides the capacity to detect
communities in different resolutions.

However, it is  expected that for networks with well-defined
community structure, the model selection criterion should be robust
to the choice of accuracy $\epsilon$. To verify this, different
accuracy $\epsilon$ ranging from $1/n$ to $1/7n$ are tested on the
journal citation network \cite{info}, this criterion identifies 4
communities for $\epsilon$ ranging from $1/n$ to $1/6n$ and 3
communities when $1/7n$, strongly indicating that this network
actually has 4 communities. We further test how different $\epsilon$
will impact on the model selection result using the Symmetric Test
when $k_{out}=6,7,8$, respectively. For $\epsilon$ ranging from
$1/2n$ to $1/7n$, this criterion nearly always identifies the
correct number of communities when $k_{out}=6,7$, however, when
$k_{out}=8$, the accuracy drops drastically, this is due to the
fuzzy structure when there are too many edges linking to other
communities. The above results shows that the model selection
criterion for \textbf{SPAEM} indeed is robust to choice of
$\epsilon$ for well clustered networks.

\section{CONCLUSION}
In this paper, we propose a probabilistic algorithm \textbf{SPAEM}
to resolve community structure. We have showed the power of
\textbf{SPAEM} in detecting community structure as well as providing
more useful information. \textbf{SPAEM} is also extended to handle
weighted network. To determine the optimal number of communities,
minimum description length principle is employed and tested on a
variety of networks.

The mixture model in \cite{newmanEM} is  a good algorithm capable of
detecting patterns and handling directed networks while
\textbf{SPAEM} focuses on detecting community structure.
Experimentally \textbf{SPAEM} does perform better in uncovering
community structure and identifying overlapping nodes. Though these
two algorithms seem to be similar with each other, they are based on
different model assumptions. Table \ref{table:cmp} gives a summary
on features of the two algorithms.
\begin{table}
\caption{\label{table:cmp} Summary table: features of SPAEM and the
Mixture model \cite{newmanEM}}
\begin{ruledtabular}
\begin{tabular}{lcc}
 & SPAEM & Mixture
\\
\hline
 Time Cost & $O(cl)$ & $O(cl)$
 \\
  Model Selection? & Yes & No
 \\
 Weighted Graph?& Yes & No
 \\
 Directed Graph?& No & Yes
 \\
 Detect Pattern?& No & Yes
\end{tabular}
\end{ruledtabular}
\end{table}
\section{ACKNOWLEDGMENT}
This paper is supported by Science Fund for Creative Research Group,
Chinese Academy of Science, No 10531070. The authors thank Dr.
Martin Rosvall in Washington University and Dr. Lingyun Wu and
Shihua Zhang in Chinese Academy of Science for their insightful
comments on this paper. We specially thank the two anonymous
reviewers for their review comments which help us to further explore
the feature of our algorithm.

\end{document}